\documentclass{iopart} % final
\usepackage{iopams} 

\usepackage{verbatim}
\usepackage{pgf}
\usepackage{tikz}

\usepackage{graphicx}

\usepackage{color}

\usepackage[normalem]{ulem} % package to 'strike out' text with \sout{bla}

\begin{document}

\title[Triple Michelson for a Third-Generation Gravitational Wave Detector] 
{Triple Michelson Interferometer for a Third-Generation Gravitational Wave Detector}

%\author{A~Freise$^1$, S~Chelkowski $^1$, S~Hild$^1$, W~Del~Pozzo$^1$, A~Perreca$^1$, A~Vecchio$^1$} 
%\address{$^1$ School of Physics and
%Astronomy, University of Birmingham, Edgbaston, Birmingham B15 2TT, UK}
\author{A~Freise, S~Chelkowski, S~Hild, W~Del~Pozzo, A~Perreca and A~Vecchio} 
\address{School of Physics and
Astronomy, University of Birmingham, Edgbaston, Birmingham B15 2TT, UK}

\ead{adf@star.sr.bham.ac.uk}

\begin{abstract}
The upcoming European design study  `Einstein gravitational-wave Telescope' 
represents the first step towards a substantial, international effort for the 
design of a third-generation interferometric gravitational wave detector.
It is generally believed that third-generation instruments
might not be installed into existing infrastructures but will provoke a new 
search for optimal detector
sites. Consequently, the detector design could be subject to 
fewer constraints than the on-going
design of the second generation instruments. In particular, it 
will be prudent to investigate alternatives to the 
traditional {\sf L}-shaped Michelson interferometer. In this article, 
we review an old proposal to use 
three Michelson interferometers in a triangular configuration. 
We use this example of a triple
Michelson interferometer to clarify the terminology and will put 
this idea into the context of more 
recent research on interferometer technologies.
Furthermore the benefits of a triangular detector will be used to motivate 
this design as a good starting point for a more detailed research effort towards 
a third-generation gravitational wave detector.
\end{abstract}

\pacs{04.80.Nn, 07.60.Ly, 95.75.Kk, 95.55.Ym}

% 04.80.Nn Gravitational wave detectors and experiments 
% (see also 95.55.Ym Gravitational radiation detectors; 
% mass spectrometers; and otherr instrumentation and techniques) 
% 07.60.-j Optical instruments and equipment (see also 87.64.M-
% Optical microscopy in biological and medical physics) 
% 07.60.Ly Interferometers % 95.75.-z Observation and data reduction techniques; 
% computer modeling and simulation % 95.75.Kk Interferometry 
% 95.55.-n Astronomical and space-research instrumentation (see also 94.80.+g 
% Instrumentation  for space plasma physics, ionosphere, and magnetosphere) 
% 95.55.Ym Gravitational radiation detectors; mass spectrometers; and other 
% instrumentation  and techniques (see also 04.80.Nn Gravitational wave 
% detectors and experiments in-General % relativity and gravitation)

\section{Introduction}

The first generation of large-scale laser-interferometric gravitational-wave
detectors, consisting of GEO\,600 \cite{geo}, Virgo \cite{virgo}, LIGO
\cite{ligo} and TAMA300 \cite{tama} is now in operation and collects data of
unprecedented sensitivity and bandwidth. All of these detectors have successfully
performed long-duration data recording runs. The path for the second generation
of laser-interferometric detectors is clearly laid out: strong R\&D projects are
currently being carried out for Advanced LIGO \cite{advligo}, Advanced Virgo
\cite{advvirgo}, LCGT \cite{LCGT} and GEO-HF \cite{geohf}, of which all except
the LCGT project are planned as advanced technology upgrades of the existing detectors.
\enlargethispage*{\baselineskip}

However, these second generation detectors are expected to approach the sensitivity limits 
given by the current infrastructures. The use of new detector sites can be an interesting 
alternative. Such sites, especially if underground, could not only allow third-generation 
detectors to improve the sensitivity by a factor of ten in
a wide frequency range but would be an investment providing enough scope for future
upgrades of the instrument over a substantial period of time.
% so that a further increase in sensitivity would require
%new facilities. 
In Europe a broad collaboration, including the GEO and Virgo
groups has begun a design study for a third-generation gravitational-wave
detector called `Einstein gravitational-wave Telescope' (ET) 
\cite{EThal, ETdesign}. 
This project aims at building an instrument that
provides
%The challenging goal of this project is to build an instrument providing 
a strain sensitivity about a hundred times better than first generation
detectors\footnote{This is equivalent to a strain sensitivity ten times better than
those planned for second-generation detectors.} and to shift the lower end of the
observational window to frequencies of approximately $1$\,Hz~\cite{Hild08}.

This article introduces some useful concepts and methods for 
the classification of possible designs of third-generation detectors. 
In Section~\ref{sec:geometry} we
discuss the concepts of detector \emph{geometry}, \emph{topology}
and \emph{configuration}.
%, of which the topology describes the general
%shape and of the core interferometer. 
In the following Sections 
%we are presenting a review of research on interferometer topologies and its relevance
%to third generation ground-based detectors. In particular 
we review
aspects of triangular detector topologies which, to some extent, have been
discussed within the context of the space-borne detector 
LISA~\cite{LISAweb}. %,LISA, LISASYMP, LISAcode}.
We recall an early proposal of a ground-based, triangular set of co-located 
Michelson interferometers originally suggested by R\"{u}diger, 
Winkler and collaborators~\cite{MPQ-talk, MPQ-report} and provide further
arguments for this geometry and topology.
While this design represents only one of many possibilities for future
detectors, it features the best understood
long-baseline interferometer, the Michelson interferometer, yet offers
interesting opportunities for \emph{virtual interferometry} 
(see Section~\ref{sec:virtual}).
In particular, it allows the construction of a simple \emph{null-stream} 
(see Section~\ref{sec:multimi}),
while being cost-effective to build (see Section~\ref{sec:tubes}).
%In section~\ref{sec:polarisation} we
%show that a triangular Michelson topology can detect and discriminate
%gravitational waves of both polarisation at any time. 
We conclude in Section~\ref{sec:summary} with a summary and outlook.
\begin{figure}[Htb] 
\centering 
\includegraphics[width=.7\textwidth]{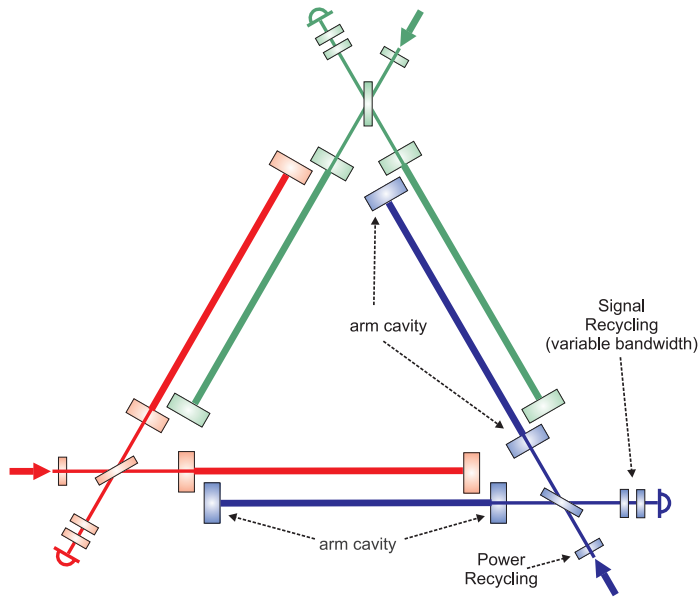}
\caption{Example optical 
layout of an interferometric gravitational-wave detector 
with a triangular geometry. The sketch shows a detector 
formed by three coplanar interferometers that form an equilateral triangle. The
interferometers are based on the Michelson topology. However, the interferometer
configuration includes additional optical technologies, like resonant arm cavities, power recycling, 
and a tunable cavity as a signal recycling mirror. Further information about the detailed
detector configuration would be provided by the parameters of the optical elements and the
operating point of the mirror position control system. This example shows how three 
state-of-the-art
interferometers could be combined to a new type of detector.} 
\label{fig:triplemi} 
\end{figure}

%%%%%%%%%%%%%%%%%%%%%%%%%%%%%%%%%%%%%
%%%%%%%%%%%%%%%%%%%%%%%%%%%%%%%%%%%%%
%%%%%%%%%%%%%%%%%%%%%%%%%%%%%%%%%%%%%

\section{Detector Geometry, Topology and Configuration}\label{sec:geometry}
The terms geometry, topology and configuration are often used loosely to describe the 
type of an interferometer, its optical layout or its physical dimensions. 
We propose to use the following
definitions for describing the location, type and optical layout of interferometric
detectors. 
%Each definition refers to a single detector site. 
\begin{itemize}
\item \emph{Geometry}: This describes the position information of  one
or several interferometers,
defined by the number of interferometers, their location and
relative orientation. 
The detector shown in 
Figure~\ref{fig:triplemi} is defined by the following geometry: three 
interferometers of equal arm length are located in a plane. The interferometer arms are 
aligned such that  together they form an equilateral triangle.
\item \emph{Topology}: 
The topology describes the optical system formed by its core elements,
examples are the classical Michelson, Sagnac and Mach-Zehnder topologies~\cite{Hecht}.
The triple Michelson illustrated in Figure~\ref{fig:triplemi} 
utilises the Michelson topology, even though it employs
arm cavities and recycling techniques. 
\item \emph{Configuration}: This describes the detail of the optical layout and
the set of parameters that can be changed
for a given topology, ranging from the specifications of the 
optical core elements to the control systems, including the operation 
point of the main interferometer. Also the addition of optical
components to a given topology is often referred to as a change in configuration.
\end{itemize} 

In order to reach their ambitious goals, third-generation detectors will very likely be located
deep underground. First of all, this can significantly reduce seismic noise and
gravity gradient noise (see \cite{Hughes98, Thorne99} for a review on gravity
gradient noise and \cite{CLIO03} for a comparison between the seismic noise at
the underground and surface interferometers in Japan). Secondly, going
underground might provide a relatively easy realization of (very tall) low
frequency suspensions. Regardless of the details of the implementation, it is
clear that a new infrastructure will allow us to design interferometers that
are completely different from the single Michelson that characterises present
laser-interferometric detectors. Hence the interferometer geometry and topology
become an important area of research\footnote{The ET design study proposal emphasises
the option for an instrument based on a new topology.}.

For the full extraction of astrophysical information, and in particular the source position 
in the sky (for short-lived signals as those from {\em e.g.} coalescing binaries and supernovae) 
a network of largely separated instruments is mandatory. The design and geometry of such a 
detector network is not the subject
of this article. Instead we concentrate on the interferometer geometry at
one detector site. In particular, we describe the benefits of using three Michelson
interferometers in one location, using a triangular geometry.
%An example optical layout of such a detector is depicted in Figure~\ref{fig:triplemi}.

The design of a third-generation interferometer will 
probably take place in two phases. During an initial phase the analysis of 
advantages and disadvantages of different geometries, topologies and 
configurations can be pursued independently. During the second phase, 
however, a system design, including all aspects of the interferometer, 
will be required. 
The remainder of the paper focuses on the first 
phase and investigates the merits of a triangular geometry;
we will review some features of single and multiple
Michelson interferometers to conclude that a triangular set of three Michelson 
interferometers in a plane combines the most interesting features.

%%%%%%%%%%%%%%%%%%%%%%%%%%%%%%%%%%%%%
%%%%%%%%%%%%%%%%%%%%%%%%%%%%%%%%%%%%%
%%%%%%%%%%%%%%%%%%%%%%%%%%%%%%%%%%%%%

\section{Virtual Interferometry}\label{sec:virtual}
The term \emph{virtual interferometry} is used in the literature for describing
various techniques. Most commonly it describes the use of numerical simulations
to study the features of an interferometer \cite{Brillet03} or techniques in
astronomy where an interference between two measured optical signals is
performed numerically as part of the data analysis process \cite{Cai06}. We
propose to use the term with respect to interferometric gravitational-wave
detectors for describing linear combinations of interferometer output 
signals that provide a readout equivalent to an additional optical interference.

The current literature on gravitational wave detection shows two very
interesting and as yet unrelated methods of combining interferometer signals
numerically for enhancing the sensitivity of a detector. The 
most prominent example is the  technique of \emph{time-delay interferometry} (TDI;
see~\cite{DhurandharTinto2005} for a review, and references therein), a
technique developed primarily to suppress the otherwise overwhelming
contribution from laser frequency noise in LISA. It creates the main detector output
signals (also called observables in this context) by time shifting and linearly
combining the many available interferometer output 
signals~\cite{NayakVinet2004}.%,Vallisneri2005}. 
These observables
can be understood as the output of `virtual' interferometers. Examples
are the two $60^\circ$ pseudo-Michelson observables with uncorrelated
noise~\cite{Cutler98} and a so-called \emph{Sagnac}
observable~\cite{Tintoetal2001}, in which the contribution from gravitational
waves is largely suppressed at frequencies lower than the inverse of the
round-trip light-time of photons along the arms of the instrument.

The second prominent application of numerical combinations of optical signals is
called \emph{displacement-noise free interferometry}~
\cite{Chen06b,Chen06a}. 
%\cite{Chen06b,Kawamura04,Chen06a,Sato07}. 
Even though it uses a very similar idea of combining 
interferometer signals that contain the gravitational wave signal with different 
phase information, the aim is slightly different. This method, so
far exclusively aimed at ground-based detectors, can be used to discriminate
gravitational radiation from signals that are generated by any other
displacement of the main mirrors, for example, through seismic or thermal noise.
The simple term `displacement noise' covers many limiting noise sources
of state-of-the art detectors. Removing or reducing such noise would be equivalent
to an improvement in sensitivity over a wide frequency range, which so far no other proposed
technology can promise. The proposed displacement-noise-free methods
are so far still proof-of-principle designs and far from any practical application. Moreover,
their effective noise reduction is most effective at frequencies
above $\sim c/L$~\cite{Chen06a}.
For ground-based detectors, of which even 
future detectors are likely to have arm lengths less than 30\,km, this frequency region
$f>10$\,kHz is typically not dominated by displacement noise but shotnoise.
%$f>10$\,kHz is not of principal interest\footnote{We note that
%this frequency range is still of considerable importance for the study of 
%neutron star normal modes, see \emph{e.g.}~\cite{AnderssonKokkotas:1998} and 
%references therein.}.
However, research into these
technologies is on-going; the potential for increasing the detector sensitivity by a large 
factor and over a wide band makes displacement-noise reduction one of the most 
exciting ideas for future interferometers.
\enlargethispage*{2\baselineskip}

Both TDI and displacement-noise free interferometry utilise extra
interferometer output channels to dramatically improve the performance
of the respective instrument.
Future detectors will probably make increasing use of  such `virtual
interferometry'. Regardless of the exact implementation, this requires multiple
readout channels for the gravitational-wave signal, which can be achieved
by using co-located interferometers or alternatively by implementing new interferometer
configurations. Both methods have already been used to create so-called null-stream channels
(see below).

%%%%%%%%%%%%%%%%%%%%%%%%%%%%%%%%%%%%%
%%%%%%%%%%%%%%%%%%%%%%%%%%%%%%%%%%%%%
%%%%%%%%%%%%%%%%%%%%%%%%%%%%%%%%%%%%%
\section{Advantages of Multiple co-located Interferometers}\label{sec:multimi}
\subsection{Generation of Null-Streams}
An important part of the data analysis for gravitational-wave detectors is the
early identification of `false candidates'; noise events that could be
mistaken for gravitational-wave signals. A powerful method, especially 
regarding unmodelled sources, is the construction of so-called
\emph{null-streams}. A null-stream is defined as a data stream formed by 
a linear combination of detector signals such that the gravitational-wave contribution 
exactly cancels  (within the calibration accuracy), while signals of other origin are left with a finite amplitude.
Therefore every event that can be detected in the null-stream as well as the
standard detector output can be discarded as noise.

The GEO\,600 detector employs
a null-stream which is created from two output channels of the main
interferometer, both of which carry the gravitational wave signals
\cite{geonullstream}.  This type of null-stream is aimed mainly at
identifying noise events that originate in the data acquisition
system. In addition, it recognises disturbances inside the
interferometer that have an optical transfer function different from
gravitational-wave events. However, this particular type of
null-stream technique cannot distinguish between gravitational waves
and similar optical signals like those from displacements of the
interferometer optics.

Null-streams in the context of burst analysis for a network of
detectors were first investigated by G\"{u}rsel and
Tinto~\cite{nullstream}. More recently, this technique has received
much attention and have been further
developed~\cite{h1h2,Chatterji06,nullstream-wen} due to the
availability of large data sets collected during science runs and
the need to employ robust methods to discriminate between transient
noise fluctuations and signals of astrophysical origin.

In general it is always possible to form a null-stream from three, not all
co-aligned, detectors. The sensitivity of this stream to noise events
does however depend strongly on the relative instrument rotation.
%In general, given three instruments (not all co-aligned) it is always
%possible to form one null-stream albeit the sensitivity of 
%this stream to noise events depends strongly on the relative
%instrument rotation. 
A general formalism to construct null-streams 
was developed in~\cite{Chatterji06} and 
this technique is used in a number of on-going searches.  
If the instruments are aligned,
two interferometers allow us to synthesise an observable which is insensitive
to gravitational waves~\cite{Chatterji06}. The simplest null-stream
can be created in real time from a pair of redundant detectors as done
at LIGO-Hanford~\cite{h1h2}:
Two co-aligned Michelson interferometers have been installed on the same site
so that the null-stream that cancels gravitational waves of all kinds at all
times can be computed by simply taking the difference of the 
detector signals:
\begin{equation}
h_{\rm null}=h_1-h_2\,,
\end{equation}
where $h_i$ stands for the main detector channel calibrated in gravitational-wave strain.

\begin{figure}[Htb] 
\centering 
\includegraphics[width=.8\textwidth]{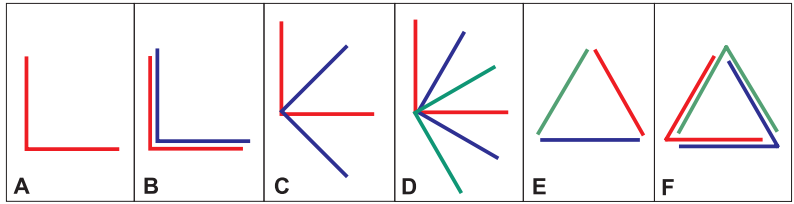}
\caption{A comparison of several geometries for future ground-based detectors: 
\textbf{A:} A simple Michelson interferometer is sensitive only to a linear
combination of the two polarisation amplitudes.
\textbf{B:} Two co-aligned Michelson interferometers 
provide redundancy and the possibility to generate a null-stream (and as for case A are sensitive only to a linear
combination of the two polarisation amplitudes).
\textbf{C:} Two Michelson interferometers rotated $45^\circ$
with respect to each other can fully resolve both polarisation amplitudes.
\textbf{D:} Three rotated Michelson interferometers provide
redundancy and the possibility to generate a null-stream. They also can measure
both polarisations (the geometries shown as C and D feature
intersection tubes. Similar geometries in which the Michelson interferometers do not overlap
might be more practical, depending on the properties of the detector site, see \cite{MPQ-report}). 
\textbf{E:} A
LISA-like triangular configuration, in which the interferometer arms are single
cavities and there is no optical recombination. 
\textbf{F:} A Triple Michelson interferometer
configuration consisting of three individual Michelson interferometers.
%Both of the latter configurations are redundant and measure both
%polarisations as configuration C. Due to the $60^\circ$ angles of the arms the overall sensitivity is
%decreased by 13\,\% in amplitude, however these configurations require only half of the tunnel length of 
%configuration D and use only three end stations. This results in a greatly decreased installation
%cost.
} 
\label{fig:configs} 
\end{figure}

An equally simple and powerful null-stream can be obtained using
three Michelson interferometers located in a plane. Using the
framework developed in \cite{nullstream-wen, jaranowski} it can be
shown that three Michelson interferometers orientated at $0^\circ$,
$30^\circ$ and $60^\circ$ (see Figure\,\ref{fig:configs}\,D)
represent a fully redundant set such that the output signal of each
individual interferometer can be simply generated from the
respective other two without undue amplification of the detector noise.
In particular, for three Michelson interferometers
oriented at $0^\circ$, $120^\circ$ and $240^\circ$, one obtains:
\begin{equation}
-h_{0^\circ}= h_{240^\circ}+h_{120^\circ}\,,
\end{equation}
where the sign of the operation is defined by which ports of the Michelson interferometers
are used to inject the laser light (see \ref{sec:math} for a brief derivation
of this relation). It follows that using this detector geometry a null-stream can 
again be created by a simple addition of the three main interferometer outputs.

This shows that a detector composed  of
three Michelson interferometers, which are rotated by $30^\circ$ or $120^\circ$ 
with respect to each other, features redundancy and null-stream
capabilities like those of co-linear Michelson
interferometers. Furthermore, this set of three Michelson interferometers 
provides redundancy to maintain full sensitivity to \emph{both} gravitational
wave polarisations, see next section.

%%%%%%%%%%%%%%%%%%%%%%%%%%%%%%%%%%%%%
%%%%%%%%%%%%%%%%%%%%%%%%%%%%%%%%%%%%%
%%%%%%%%%%%%%%%%%%%%%%%%%%%%%%%%%%%%%
\begin{figure}[tbh] 
\centering 
\includegraphics[viewport=0 30 860 370,width=\textwidth]{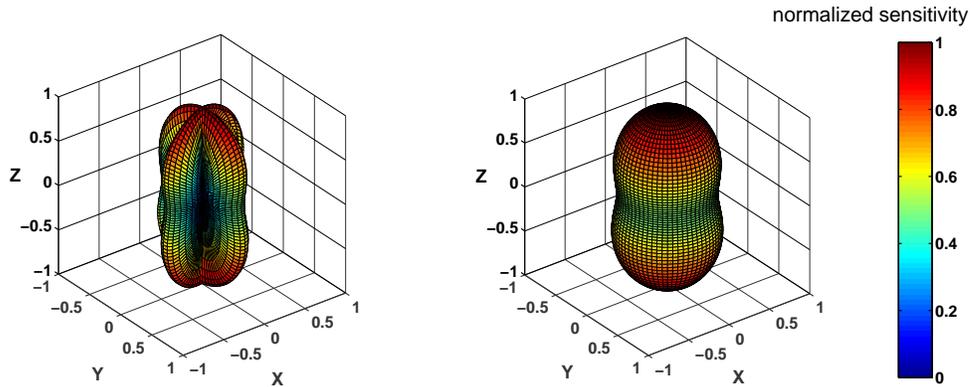} 
\caption{The response of a detector to a linear polarised gravitational wave
as a function of the detector orientation. Both plots show the normalised sensitivity
to a wave travelling along the z-axis. Each data point represents the sensitivity of the
detector for a specific detector orientation defined by the detector normal
passing the respective data point and the origin. The colour
of the data point as well as its distance from the origin indicate the magnitude of the
sensitivity. The left plot depicts the response of a single Michelson, while the right plot
gives the response of a set of three Michelson interferometers in a triangular geometry
as shown in Figure\,\ref{fig:configs}\,F.} \label{fig:pol} 
\end{figure}

\subsection{Sensitivity to the Gravitational-Wave Polarisation}\label{sec:polarisation}
A Michelson interferometer provides maximal sensitivity to a specific
polarisation (or, equivalently measures only a linear combination of the two polarisation amplitudes). 
Two interferometers rotated by $45^\circ$ as shown in Figure\,\ref{fig:configs}\,C allow full
reconstruction of both polarisation amplitudes.
Generally, gravitational 
waves will not arrive at the detector
in the optimal polarisation for one interferometer, thus the detection of the second 
polarisation increases the total signal strength (see Figure~\ref{fig:pol}).
Furthermore, 
two such oriented Michelson interferometers can be used to detect both polarisations
and to determine the polarisation angle. This is important to fully resolve the geometry of
a source and test general relativity~\cite{Will:2006}.

It is straightforward to show that three 
Michelson interferometers, which are rotated with respect to each other in an
appropriate way, can also measure the polarisation of the gravitational 
wave. By the same analysis as for the null-stream 
construction it can be shown that 
we can synthesise  the output signal of a `virtual' Michelson\footnote{Note that this is equivalent to 
the well-know result derived for the LISA detector~\cite{Cutler98}.}
at $45^\circ$ from two otherwise equal interferometers rotated by 
$120^\circ$  and $240^\circ$:
\begin{equation}
h_{45^\circ}=\frac{1}{\sqrt{3}}\left(h_{240^\circ}-h_{120^\circ}\right) 
\end{equation}
Therefore the set of three Michelson interferometers offers the same advantages with respect
to gravitational wave polarisation as two interferometers that are oriented by
$45^\circ$ to each other.

So far we have used the term redundancy as an equivalent to null-stream generation.
However, we should consider redundancy also under operational aspects. The 
fact that we reconstruct a third Michelson interferometer from two other means
that we can perform hardware upgrades or maintenance sequentially on the entire
detector without interrupting data taking. In fact, with one Michelson not
operating we would still be able to fully detect both gravitational wave
polarisations, and only the construction of a null-stream would become impossible.
The possibility of having a duty cycle as close as possible to 100\% 
for the gravitational wave data channels 
becomes an important asset, especially when we consider each detector to be 
part of a larger network.

%%%%%%%%%%%%%%%%%%%%%%%%%%%%%%%%%%%%%
%%%%%%%%%%%%%%%%%%%%%%%%%%%%%%%%%%%%%
%%%%%%%%%%%%%%%%%%%%%%%%%%%%%%%%%%%%%
\section{Interferometer Topologies }\label{sec:topologies} 
%The typical {\sf L}-shape of current detectors does not necessarily require a Michelson 
%interferometer but can be achieved with other interferometer types, as shown in 
To date no laser interferometer topology other than the Michelson has been
used for gravitational wave detection. However, some very advanced
noise reduction techniques proposed for future 
detectors are based on topologies of the Sagnac interferometer, the Fox-Smith cavity or the
Mach-Zehnder interferometer~\cite{Chen03,danilishin2006,Chen06b}.

\begin{figure}[thb] 
\centering 
\includegraphics[width=\textwidth]{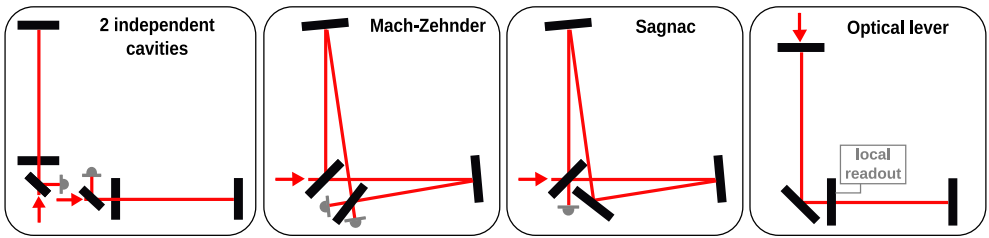}
\caption{Four example interferometer topologies that can be used in a {\sf L}-shaped form: 1) two single cavities
2) Mach-Zehnder interferometer 3) Sagnac interferometer 4) Optical lever.}
\label{fig:Lshapes} 
\end{figure}

It is worth noting that a triangular geometry as discussed above
is conceivable with different interferometer
topologies. In particular it is possible to use different
topologies while maintaining the {\sf L}-shape of the single
interferometers
as displayed in figure~\ref{fig:Lshapes}.
Therefore, for example, three Sagnac interferometers or three cavities
could be used to form a triangle.
Such detector designs can provide similar benefits as described above for the 
triple Michelson geometry so that the triangular geometry is largely independent 
of the topology of the individual interferometers. 
%However, each interferometer topology offers various
%advantages and disadvantages regarding a third generation gravitational wave detector 
%and each should be re-evaluated in this context. 

The case for alternative topologies is largely based on ideas for the
reduction of quantum noise. In general, the signal-to-noise ratio of a 
single interferometer is different for each topology, with the actual
difference depending also on the type of noise under investigation.
However, it is not possible to identify a topology with a meaningful 
signal-to-noise ratio or sensitivity since these vary dramatically
with the interferometer \emph{configuration}. Consequently, detailed
interferometer designs must be studied for comparing different 
topologies. To-date such effort has only been fully undertaken for 
the Michelson topology including ongoing research which shows that the Michelson 
topology offers interesting possibilities for new 
quantum noise reduction techniques~\cite{rehbein08}.

During the design and construction of the first generation of detectors 
the Sagnac topology has been investigated and prototypes have been
build~\cite{sun96} but eventually it did not show significant advantages over the Michelson 
topology~\cite{Mizuno97}. More recently it has been proposed to use the Sagnac topology 
as a \emph{speed meter}~\cite{Chen03} to reduce the quantum noise.
The Sagnac topology can be hosted in different ways in a triangular
geometry: each Sagnac as an equilateral triangle, or as an {\sf L}-shaped
zero-area Sagnac. Noise couplings due to the Sagnac effect favor the 
zero-area Sagnac topology: it can be shown that for a typical
choice of optical parameters this extra noise couplings do not
impose stringent new requirements in the case of a zero-area Sagnac
interferometer, see~\ref{sec:sagnac}.

A detailed study of the proposed alternative topologies will be a
core activity of the ET design study and is beyond the scope of this
paper. We note that Michelson-based detectors currently offer the 
advantage of using the experience as well as the 
advanced optical technologies of the first two detector generations
and thus at least must be taken as a reference against which other topologies must
be compared.

\section{Triple Michelson Interferometer}\label{sec:tubes} 
A triangular detector geometry as
displayed in Figure\,\ref{fig:configs}\,F has been proposed already in 1985 
by R\"{u}diger, Winkler and collaborators~\cite{MPQ-talk, MPQ-report}. 
Henceforth we will call this layout a {\em triple Michelson} to differentiate 
this geometry from other triangular ones.
%Table~\ref{tab:summary} compares the features of the triple Michelson with those 
%of the other geometries shown in Figure~\ref{fig:configs}. 

%Furthermore, the 
%experimental work with the first and soon the second generation of interferometric
%detectors has created vast amounts of experience with this interferometer type.

It is useful to consider whether a triangular configuration would
allow us to reconstruct completely the geometry of an arbitrary
gravitational wave source. For simplicity we consider the case of a
(non-spinning) binary system, but the counting argument that we
present here equally applies to other burst signals 
(it is not appropriate however for stochastic signals or long-lived 
signals, such as those from rotating neutron stars).
%or continuous
%waves from rotating neutron stars (it is not appropriate however for
%stochastic signals). 
In order to reconstruct the geometry of a source
one needs to estimate (at least) five parameters: the luminosity
distance to the source, two angles that identify the source position
in the sky, and two angles that describe the orientation of the
orbital angular momentum (one usually uses the polarisation angle
$\psi$ and the inclination angle $\iota$). A single Michelson
interferometer would provide only two independent quantities, and the
problem would be severely underdetermined. A triangular topology, on
the other hand would provide \emph{four} independent constraints to
measure five unknowns. An additional  interferometer in a different
location would therefore be still necessary to fully break the
degeneracy, which is however much less severe than in the case of a
single Michelson. 

The triangular geometry is equivalent to the
one in Figure\,\ref{fig:configs}\,D, with the only difference\footnote{The 
three Michelsons are not exactly co-located, for 10\,km 
long arms the interferometer signals can be subject to a time delay 
up to the order of $L/c\approx0.3$\,ms.} that the $60^\circ$
opening angle of the interferometers reduces the strain sensitivity to
$\sin(60^\circ) = 0.87$ of the optimal one. Such a moderate
loss of sensitivity is compensated by 
a substantial reduction in the required underground space:
compared with the geometry depicted in Figure\,\ref{fig:configs}\,D, a triangular
detector only requires half the tunnel length and only 3 instead of 7 end
stations. 
%As we have shown a triangular configuration is equivalent to two
%Michelson interferometers with 60$^\circ$ opening angle and rotated
%one with respect to the other by 45$^\circ$. 
In can also be shown that the sensitivity of the triple Michelson is
very similar to a set of four right-angled Michelson detectors oriented at
$0^\circ$ and $45^\circ$, see \ref{sec:triplevsfour}.
Such a configuration is
optimal in the sense of providing the whole number of independent
observables that can be obtained by an \emph{arbitrary} number of
co-planar and co-located instruments. In fact, the addition of 
more instruments in the same location and plane would not
change the number of independent information that can be obtained, as
one would be able to reconstruct the output of this additional
instrument as a linear combination of the readouts of the two
Michelsons given by the triangular topology. 
%Adding additional
%instruments would however increase the signal-to-noise ratio. If the
%optical layout of the triangular topology yields three instead of only
%two independent instruments, this would clearly be an advantage in
%this respect, while maintaining the total tunnel length.
The triple Michelson can be considered the minimal (in terms of
enclosed area and number of end stations) detector geometry that combines
all features of the various options using co-located, 
co-planar Michelson interferometers.

\section{Summary and Outlook} \label{sec:summary}
The Michelson interferometer is ideally suited for measuring gravitational waves. It combines
large bandwidth with good sensitivity to gravitational wave strain.
Second-generation gravitational wave detectors are based on modern Michelson interferometers 
enhanced by advanced techniques, like power- and signal recycling. 
It is expected that such detectors will beat the Standard Quantum Limit of 
interferometry in a small frequency range.~\cite{Buonanno, Chelkowski}.

Third-generation instruments might be constructed at new detector sites and
therefore new interferometer layouts can be considered. In this paper we have 
outlined the concept of detector \emph{geometry}, \emph{topology} and \emph{configuration} 
for 
investigating possible designs for third-generation detectors. 
We have further evaluated the idea of three co-located 
interferometers in a triangular geometry.
The original proposal of this geometry was motivated mainly by the idea that
such a detector had the ability to detect both gravitational wave 
polarisations. In this article we have shown that such a geometry offers
significant advantages based on `virtual interferometry' techniques.
In particular, we have highlighted the fact that the interferometer set is fully redundant 
and offers the possibility to create an efficient null-stream
from the local detector data.
Several known interferometer topologies can be employed in the triangular 
geometry and such topologies have to be re-evaluated in the light of the
demanding sensitivity of third-generation detectors. 
%We have shown that the Sagnac 
%interferometer can be constructed such that noise couplings due to the
%Sagnac effect can be negligible.

As part of the European design study 'Einstein gravitational-wave Telescope' 
a careful and in-depth system design will begin
with the aim of understanding which detector geometry and topology is
optimal for third-generation detectors.
The triple Michelson detector presents a realistic concept
as it combines a well tested and well understood interferometer
design with new possibilities, mainly based on the combination
of multiple interferometric signals. 
These advantages will play a strong role in deploying gravitational-wave
{\em telescopes} capable of continuous surveys.
Therefore this detector geometry should be considered as 
a meaningful starting point for design studies that are about to begin.

\section{Acknowledgement} We would like to thank Albrecht R\"udiger, Roland Schilling and
Harald L\"uck for many useful discussions. We are particularly grateful to  Albrecht R\"udiger
for providing us with the original proposal of the triple Michelson detector.  This work has been
supported by the Science and Technology Facilities Council (STFC)
and the European Gravitational Observatory (EGO). 
This document has been assigned the LIGO Laboratory
document number LIGO-P080019-00-Z.

\appendix
\section{Virtual Michelsons}\label{sec:math}
From \cite{jaranowski} we obtain the following general expression for the response function of a Michelson
interferometer to gravitational waves:
\begin{equation}
h(t)=F_{+}(t)h_{+}(t)+F_{\times}(t)h_{\times}(t)
\end{equation}
where $F_{+}$ and $F_{\times}$ are the beam pattern function which in turn can be written as:
\begin{equation}
\begin{array}{l}
F_{+}(t)=\sin\zeta\left(a(t)\cos2\psi +b(t)\sin2\psi\right)\\
F_{\times}(t)=\sin\zeta\left(b(t)\cos2\psi -a(t)\sin2\psi\right)
\end{array}
\end{equation}
with $\zeta$ the opening angle of the interferometer arms and $\psi$ the polarisation angle of the gravitational wave.
$a(t)$, $b(t)$ are complex functions of the detector location in space and time. We are
only interested in their dependence on the detector rotation around its normal, here described by the angle 
$\gamma$. We can thus simplify $a$, $b$ to:
\begin{equation}
\begin{array}{l}
a(\gamma)=C_1 \sin2\gamma + C_2 \cos2\gamma\\
b(\gamma)=C_3 \sin2\gamma + C_4 \cos2\gamma
\end{array}
\end{equation}
with $C_n$ as functions of time as well as the remaining parameters specifying the position of the detector
and the location of the gravitational wave source.
In the following we arbitrarily set the polarisation angle to $\psi=0$.
This yields:
\begin{equation}
\begin{array}{cl}
h(\gamma)=&\sin\zeta\left[\left(C_1\sin2\gamma+C_2\cos2\gamma\right)h_{+}\right.\\
&+\left.\left(C_3\sin2\gamma+C_4\cos2\gamma\right)h_{\times}\right]\\
\end{array}
\end{equation}
In particular we obtain:
\begin{equation}
\begin{array}{l}
h(0^\circ)=\sin\zeta\left[C_2 h_{+}+C_4 h_{\times}\right]\\
h(45^\circ)=\sin\zeta\left[C_1 h_{+}+C_3 h_{\times}\right]\\
\end{array}
\end{equation}
Using simple addition and subtraction rules for $\sin$ and $\cos$ we can further write:
\begin{equation}
\begin{array}{l}
h(120^\circ)+h(240^\circ)=-\sin\zeta\left[C_2 h_{+}+C_4 h_{\times}\right]=-h(0^\circ)\\
h(120^\circ)-h(240^\circ)=-\sqrt{3}\sin\zeta\left[C_1 h_{+}+C_3 h_{\times}\right]=-\sqrt{3}~h(45^\circ)\\
\end{array}
\end{equation}

\section{Triangle versus right angled interferometers}\label{sec:triplevsfour}
\begin{figure}[Htb] 
\centering 
\includegraphics[scale=.7]{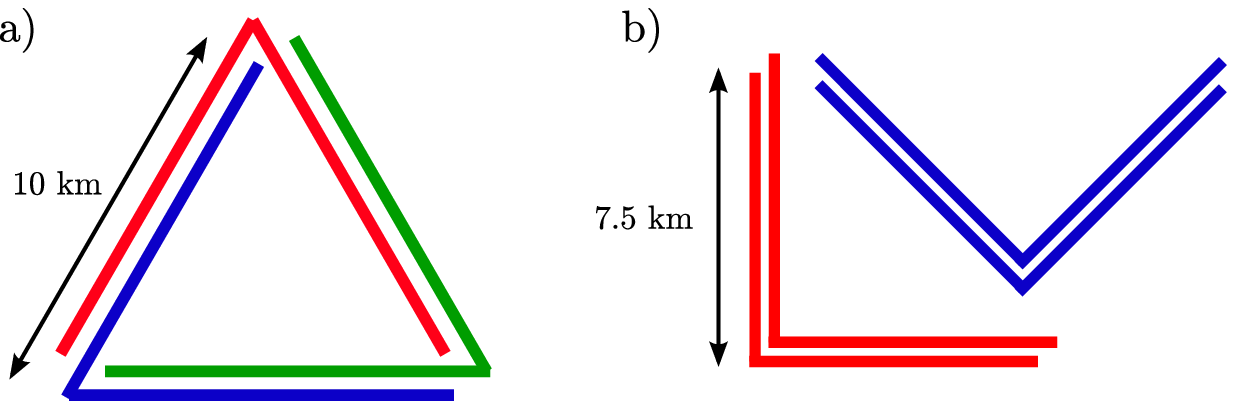}
\caption{Comparison of two detector geometries that utilise exactly the same
tunnel length and feature almost exactly the same sensitivity, see text.
The left sketch a) shows three uncorrelated Michelson interferometers
in a triangular tunnel system. On the right b) four uncorrelated 
Michelson interferometers are located in two {\sf L}-shaped tunnel
systems. In both cases the total tunnel lengths is 30\,km and each
tunnel segment hosts two interferometer arms, preferably in separate
vacuum systems.}
\label{fig:triplevsfour} 
\end{figure}
We want to compare the sensitivity of a triangular detector to that of a
similar detector that uses right-angled Michelsons. In order to do so
we assume the following constraints: both detectors should have 
the same total tunnel length and the
same number of laser beams in each tunnel segment. The two
best detector geometries within these constrains are shown in
figure~\ref{fig:triplevsfour}.
In addition, we consider all interferometers of one detector to be uncorrelated
(probably housed in separate vacuum systems). 

It is useful for this example to relate the SNR of any given 
detector to a reference instrument, here, a single Michelson interferometer
with orthogonal arms of 10\,km length. The signal strength of the triangular setup 
shown in figure~\ref{fig:triplevsfour} can be compared to the reference
detector by combining the three interferometer signal. For $\gamma=0^\circ$ we obtain:
\begin{equation}
h_{\rm \Delta}(0^\circ)=h(0^\circ)-h(120^\circ)-h(240^\circ)=2h(0^\circ)
\end{equation}
With $h_\Delta$ the signal output of the triangular detector. We can then approximate:
\begin{equation}
\frac{{\rm SNR}_{\Delta,L=10\,{\rm km}}}{{\rm SNR}_{\rm MI, L=10\,{\rm km}}}=\frac{2}{\sqrt{3}}\,\sin(60^\circ)=1
\end{equation}
And for $\gamma=45^\circ$ we obtain:
\begin{equation}
\frac{{\rm SNR}_{\Delta,L=10\,{\rm km}}}{{\rm SNR}_{\rm MI, L=10\,{\rm km}}}=\frac{\sqrt{3}}{\sqrt{2}}\,\sin(60^\circ)\approx 1.06
\end{equation}
Similarly we can compute the SNR of the right-angled setup b) by adding the 
signals of two parallel interferometers (oriented either at $0^\circ$ or $45^\circ$):
\begin{equation}
\frac{{\rm SNR}_{\rm two\,MI, L=7.5\,{\rm km}}}{{\rm SNR}_{\rm MI, L=10\,{\rm km}}}=\frac{2}{\sqrt{2}}~\frac{7.5}{10}\approx1.06
\end{equation}
This shows that the triangle geometry
%with only three corner stations and a smaller enclosed area 
has in the worst case a 6\% lower sensitivity. However, in practice, the sensitivity of the right-angled detector
might be worse since the condition of uncorrelated
noise will be much more difficult to achieve because always two interferometers share exactly 
the same location. 

\section{The Sagnac effect}\label{sec:sagnac}
\begin{figure}[htb] 
\centering 
\includegraphics[clip,width=\textwidth]{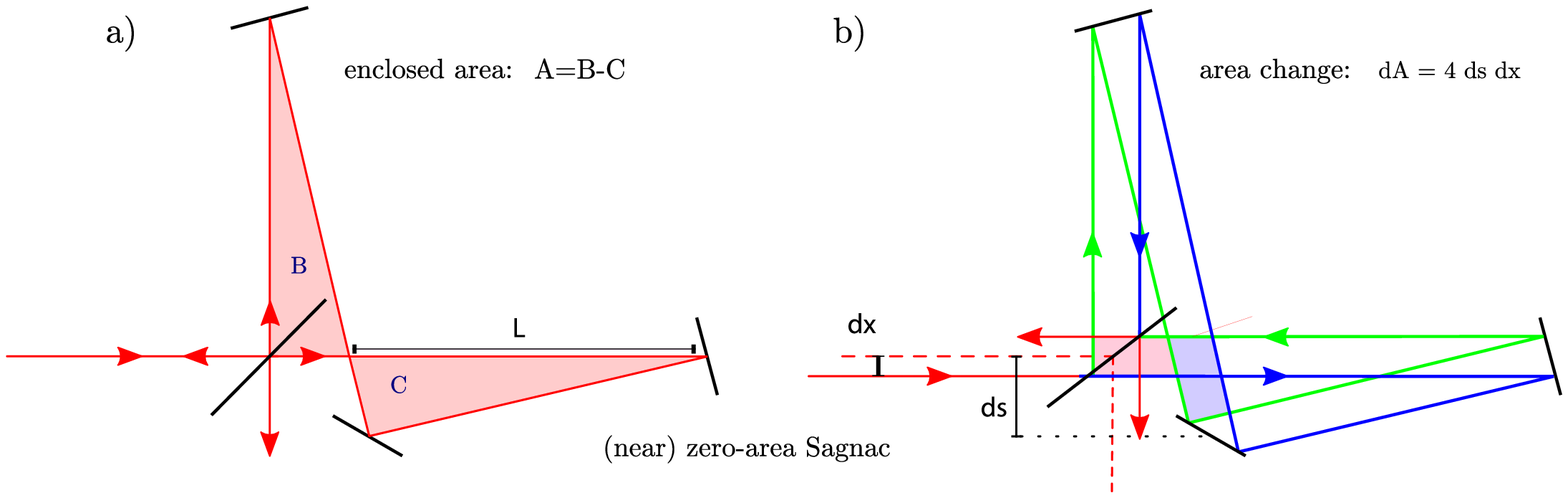} 
\caption{
a) Zero-area Sagnac interferometer: by carefully adjusting the arm length and the angle
of the end mirrors the enclosed areas $B$ and $C$ are made equal such that the
effective enclosed area $A$ is zero. b) Area change due to transverse beam displacement: in practice
a small change of the beam position, or the interferometer alignment will cause the enclosed area
to change, so that $A$ is not zero on average and also fluctuating at higher frequencies.
}
\label{fig:sagnac1} 
\end{figure}

%One disadvantage of the Sagnac topology is the extra coupling
%of noise due to the Sagnac effect.
%They might even turn out to be more advantageous for implementing advanced
%noise reduction techniques: The Sagnac interferometer, for example, might be used as a
%\emph{speed meter}~\cite{Chen03}, whereas it is straightforward to
%provide multiple output signals using single cavities. While such ideas 
%are very interesting and need to be
%investigated, we believe that Michelson-based detectors offer the great
%advantage of using the experience as well as the advanced optical technologies
%of the first two detector generations. 
One of the prime uses of the Sagnac interferometer is to measure rotation, via the
\emph{Sagnac effect}~\cite{Malykin00}: In an otherwise undisturbed Sagnac interferometer
the relative phase of the beams interfering
at the beam splitter is proportional to the effective enclosed area $A$, the angular 
frequency $\Omega$ of the interferometer rotation and the angular frequency of the light
$\omega$:
\begin{equation}\label{eq:sagnac}
\phi=\frac{4 A}{c^2}\Omega\omega
\end{equation}
Thus the Sagnac effect must be evaluated as a possible channel for coupling
laser frequency noise or seismic noise into the gravitational wave channel. To minimise the
effect the interferometer should be designed as a zero-area Sagnac. 
However, small misalignments of the optical beams
will cause $A$ to be non-zero, as indicated in figure~\ref{fig:sagnac1}.
In the following we present noise estimations for an example Sagnac layout
with 10\,km arm length.

At first we consider the Sagnac effect due to the Earth's rotation. We have 
arbitrarily chosen the latitude of Strasbourg and assume a constant rotation frequency
of $\Omega_S=\Omega\sin(48^\circ 33')$, where $\Omega$ is the rotational frequency of the Earth. 
Any change in the enclosed area $A$ or the laser
frequency $\nu$ will couple into the optical phase.
%\begin{equation}
%\Delta\phi=\frac{8\pi \Delta A}{c^2}\Omega\nu\quad\mbox{and}\quad\Delta\phi=\frac{8\pi A}{c^2}\Omega\Delta\nu
%\end{equation}
The latter can be interpreted as a new coupling mechanism for frequency noise with the coupling
given by the projection of the detector sensitivity to frequency noise:
\begin{equation}
\Delta\nu=\frac{L\,c^2}{A\,\Omega_S\,\lambda\,\cos(\alpha)}\cdot h\approx\frac{L\,c^2}{A\,\Omega_S\,\lambda}\cdot h
\quad\mbox{for small}~\alpha
\end{equation}
In order to estimate the enclosed area $A$ we assume an average miscentering or misalignment of the beams
to be of the order of $0.1$\,mm which corresponds to $A\approx0.5\cdot10\,{\rm km}\cdot0.1\,{\rm mm}=0.5\,{\rm m}^2$.
For an example sensitivity of $h=6\cdot10^{-24}\sqrt{\rm Hz}$, which represents the target sensitivity
of the Einstein Telescope at 10\,Hz, we can thus estimate the frequency noise
requirement to be:
\begin{equation}
\fl\Delta\nu\lesssim 1.9\cdot10^{8}\frac{\mathrm{Hz}}{\sqrt{\rm Hz}}
\left(\frac{h~ \sqrt{\mathrm{Hz}}}{6\cdot10^{-24}}\right)
\left(\frac{L}{10\,\mathrm{km}}\right)
\left(\frac{0.5\mathrm{m}^2}{A}\right)
\left(\frac{5\cdot10^{-5}\,{\rm Hz}}{\Omega_S}\right)
\left(\frac{1064\,\mathrm{nm}}{\lambda}\right)
\end{equation}
which does not cause any concern. We must also consider the coupling of seismic noise,
via two different effects. First of all, the seismic disturbances of the interferometer as a whole
includes a rotational component which can be characterised by
replacing $\Omega_S$ by an dynamic term in equation~\ref{eq:sagnac}. 
To estimate the effect we have assumed the simplest
case in which the seismic noise at the corners of the triangle is uncorrelated and we can write
the spectral density of the rotation angle as:
\begin{equation}
  \Delta\Omega_\sigma = \arctan\left(\frac{\Delta\sigma}{L}\right)
\end{equation}
with $\Delta\sigma$ the amplitude spectral density of the seismic disturbances. We can further project 
the detector sensitivity to the seismic noise level using:
\begin{equation}
  \Delta\sigma =L\, \tan\left(\frac{L\,c}{A\,\cos(\alpha)}\cdot h\right)
\end{equation}
and again we assume $\alpha\ll1$ to compute a requirement for the seismic noise:
\begin{equation}
\Delta\sigma\lesssim 1.8\cdot10^{-7}\mathrm{m}
\left(\frac{h~\sqrt{\mathrm{Hz}}}{6\cdot10^{-24}}\right)
\left(\frac{L}{10\,\mathrm{km}}\right)
\left(\frac{0.5\,\mathrm{m}^2}{A}\right)
\end{equation}

Another way in which the seismic noise can couple into optical phase noise is through misalignment
of the optics or input beam jitter. Input beam jitter can be interpreted as a change of the enclosed area
as shown in figure~\ref{fig:sagnac1}. The noise projection factor is given by:
\begin{equation}
  \Delta x =\frac{L\,c}{4\,\Omega_S\cdot ds\,\cos(\alpha)}\cdot h\\
\end{equation} 
with $ds$ being the distance between the beam splitter and the central turning mirror as shown in
figure~\ref{fig:sagnac1}. This leads to
the following requirement for input beam jitter:
\begin{equation}
\fl  \Delta x \lesssim 8.3\cdot10^{-8}\mathrm{m}
\left(\frac{h~\sqrt{\mathrm{Hz}}}{6\cdot10^{-24}}\right)
\left(\frac{L}{10\,\mathrm{km}}\right)
\left(\frac{5\cdot10^{-5}\,{\rm Hz}}{\Omega_S}\right)
\left(\frac{1\,\mathrm{m}}{ds}\right)
\end{equation}
Such a requirement is well within the performance of the existing 
mode cleaner systems of current detectors.
In summary, the noise introduced by the Sagnac effect is negligible 
provided that the interferometer is designed as a zero-area Sagnac and
that the beam alignment can be
performed accurately enough to ensure a small effective enclosed area.

\end{document}